\documentclass[preprint,5p]{elsarticle}

\usepackage{graphics,bm}
\usepackage{amsmath, amssymb}
\usepackage{epsfig}
\usepackage{graphicx}
\usepackage{amsmath}
\usepackage{wrapfig}
\usepackage{bbold}
\usepackage{color}
\usepackage[compat=1.1.0]{tikz-feynman}
\usepackage{amssymb}
\usepackage{bbold}
\usepackage{widetext}
\usepackage{mwe}
\usepackage{widetext}
\usepackage{graphicx}
\usepackage{subcaption}
\newcommand{\beq}{\begin{equation}}
\newcommand{\eeq}{\end{equation}}
\newcommand{\bqa}{\begin{eqnarray}}
\newcommand{\eqa}{\end{eqnarray}}

\def\square{\vcenter{\vbox{\hrule height.4pt
          \hbox{\vrule width.4pt height4pt
          \kern4pt\vrule width.3pt}\hrule height.4pt}}}

\newcommand{\eq}[1]{\begin{align}#1\end{align}}
\usepackage[normalem]{ulem}

\journal{Physics Letters B}

\begin{document}

\begin{frontmatter}



\title{Phases of QCD at nonzero isospin and strangeness
chemical potentials with application to pion stars}


\author{Jens O. Andersen}
\ead{andersen@tf.phys.ntnu.no}
\author{Martin Kj{\o}llesdal Johnsrud}

\ead{martkjoh@stud.ntnu.no}

\address{Department of Physics, Norwegian University of Science and Technology,
H{\o}gskoleringen 5, Trondheim, N-7491, Norway}

\begin{abstract}
We study pion and kaon condensation using three-flavor chiral perturbation theory at finite isospin  and strangeness quark chemical potentials $\mu_I$ and $\mu_S$.
The phase diagram consists of a vacuum phase and three
distinct Bose condensed phases with condensates of charged pions as well as charged and neutral kaons.
Adding electromagnetic interactions, a phase with a 
charged condensate becomes a Higgs phase 
and the resulting phase diagram is modified due to the electromagnetic mass splittings of the mesons.
The results for the pion-condensed phase are applied
to calculate mass-radius relation of pion stars.
Local electric charge neutrality is imposed by adding electrons, muons  together with their neutrionos.
Finally, we compare our results for the mass-radius relations
with those from recent lattice simulations, and find very good agreement.

\end{abstract}




\end{frontmatter}



\section{Introduction}
\label{intro}
The phase diagram of quantum chromodynamics (QCD) has received
a lot of attention since the first sketchy version of it 
appeared in the 1970s. At the time, it was expected that it consists
of two phases, a hadronic phase at low temperature and 
small baryon chemical potential $\mu_B$, and a deconfined phase at
high temperature and/or large baryon chemical potential.
In 1984, Bailin and Love~\cite{love} suggested that
QCD at large baryon chemical potential is a color superconductor.
The result is based on the fact that there is an attracting
channel in the one-gluon exchange diagram in QCD
and that perturbation QCD is valid for large $\mu_B$ due to asymptotic freedom. This
renders the Fermi surface unstable and in analogy with BCS theory leads to the formation of Cooper pairs.
Since then a number of colors-superconducting phases have
been found whose exact location is model dependent. See Refs.~\cite{raja,alford,fukurev}
for reviews.

Conventionally, the QCD phase diagram has been drawn in 
the $\mu_B$--$T$ plane, but we can add more axes to it, 
for example by allowing for an independent quark chemical 
potential $\mu_f$ for each flavor $f$. For two flavors, this
implies that we introduce an isospin chemical potential $\mu_I$
and for three flavors an additional strangeness chemical potential
$\mu_S$. The new $\mu_I$--$T$ plane has received particular attention, in part by the fact that fermion determinant in QCD at finite isospin density and zero
baryon and strangeness density is real, which implies that the system is free of the sign problem.
Thus one can carry out lattice simulations
using standard importance sampling techniques and study
pion condensation and the transition line in the
$\mu_I$--$T$ plane.
The first lattice simulations carried out more than 20
years ago~\cite{kogut1,kogut2,kogut3,kogut4,kogut5}
suggested that the phase transition
is second order with critical exponent from the
$O(2)$ universality class. At zero temperature the transition
takes place at $\mu_I=m_{\pi}$, which is expected for a
second-order phase transition.
In recent years, high precision lattice simulations have been 
carried out confirming 
the second-order nature of the transition at $T=0$.
Moreover, quark and pion condensates as functions 
of $\mu_I$ were obtained~\cite{gergy1,gergy2,gergy3,gergy4}.

Condensation of pions, kaons, and diquarks in QCD and QCD-like theories has also been studied using chiral perturbation
theory 
($\chi$PT)~\cite{isostep,isostep2,kim,kogut33,carig,carig2,usagain,mojahed,gronli}.
$\chi$PT is an effective low-energy theory based
on the global symmetries and relevant
degrees of freedom of the the fundamental theory (QCD or QCD-like)~\cite{gasser1,gasser2}.
The second-order nature of the transition at low temperatures,
where $\chi$PT is valid, as well as the
critical chemical potential being equal to the mass of the
condensing particle (at $T=0$)
were confirmed.~\footnote{This has been shown
at order ${\cal O}(p^2)$ and ${\cal O}(p^4)$ in the low-energy expansion 
of $\chi$PT and is expected to hold to all orders.}
Finally, excellent agreement between lattice and ${\cal O}(p^4)$
result in chiral perturbation for the quark and pion condensates at $T=0$
was found. A recent review of meson condensation can be found
in Ref.~\cite{mannarellirev}.

In Ref.~\cite{tomas}, the authors considered the problem of pion condensation in a
dense neutrino gas. At $T=0$, it was shown that for sufficiently large neutrino chemical potentials
and therefore neutrino densities, a charged pion condensate is formed in electrically neutral
matter in chemical equilibrium. The electron and muon provide the background electric charge
that neutralizes the charged pion condensate.
This opens up the possibility to probe the QCD phase diagram if these conditions are met, for example in the early universe. This question was addressed in Refs.~\cite{bod1,bod2,early}.
In Ref.~\cite{early}, the authors use an HRG model where the pion contribution is replaced by 
an interacting pion gas based on lattice simulations~\cite{sup}, to allow for pion condensation.
The resulting pressure is parameterized by the temperature $T$ and five different chemical
potentials, corresponding to the conserved charges $B$, $Q$, and the three lepton numbers.
Given that the baryon and electric charge per entropy ratios are tightly constrained,
the cosmic trajectories can be parameterized by the lepton asymmetries $l_{e,\mu , \tau}$.
While the sum of them is tightly constrained, the individual contributions are not.
By varying $l_e+l_{\mu}$ with $l_e=l_{\mu}$ and $l_e+l_{\mu}+l_{\tau}=0$, different
cosmic trajectories are obtained. For lepton asymmetry $l_e+l_{\mu}=0.1$ or larger,
the trajectories enter the pion condensed phase in the $\mu_Q$--$T$
plane, where the
transition line was obtained from lattice calculations~\cite{sup}.
The idea of a pion star formed by a Bose 
condensate of charged pions was first suggested in Ref.~\cite{carig2}. 
A detailed study of their properties,
with and without charge neutrality constraint, 
based on an EoS from
lattice simulations of QCD at finite
isospin density, can be found in Ref.~\cite{gergostar}.

In this paper, we continue the study of pion and kaon condensation in the context
of $\chi$PT, including electromagnetic effects. 
We apply the results for pion condensation 
to pion stars.
A detailed analysis will be presented
elsewhere~\cite{martin2}, see also~\cite{martin3}.
The letter is organized as follows. In Sec.~\ref{phases}, we
briefly discuss the chiral Lagrangian at finite chemical potentials
and map out the phase diagram at $T=0$ in the $\mu_I$--$\mu_S$
plane including electromagnetic effects. In Sec.~\ref{stars},
we discuss pion stars at finite isospin chemical potential.
We discuss the nonrelativistic and Newtonian limits as well
as the effects of imposing local charge neutrality.
The mass-radius relation is calculated and we compare
our results with those of recent lattice simulations.

\section{Phases of QCD at nonzero $\mu_I$ and $\mu_S$}
\label{phases}
The original formulation of chiral perturbation theory by Gasser and Leutwyler was in
the strong sector~\cite{gasser1,gasser2}. 
Later, electromagnetic interactions were included by 
Ecker et al.~\cite{ecker} at ${\cal O}(p^2)$, and by Urech~\cite{urech1,urech2} and by Meissner et al.~\cite{meissner}
at order ${\cal O}(p^4)$.
In the presence of electromagnetic interactions, the
${\cal O}(p^2)$ chiral Lagrangian for three-flavor QCD is
\eq{\nonumber
{\cal L}_2&=-{1\over4}F_{\mu\nu}F^{\mu\nu}+
{1\over4}f^2\langle\nabla_{\mu}\Sigma\nabla^{\mu}\Sigma^{\dagger}
\rangle
\\ &\quad
\nonumber
+{1\over4}f^2
\langle\chi^{\dagger}\Sigma+\Sigma^{\dagger}
\chi\rangle
+C\langle Q\Sigma Q\Sigma^{\dagger}\rangle
\\ &\quad
+{\cal L}_{\rm gf}+{\cal L}_{\rm ghost}
\;,
\label{lagel}
}
where $\langle A\rangle$ means the trace of the matrix $A$,
$F_{\mu\nu}$ is the electromagnetic field tensor, $\nabla_{\mu}\Sigma=\partial_{\mu}\Sigma-i[v_{\mu}-A_{\mu}Q,\Sigma]$  is the covariant derivative, $\chi=2B_0{\rm diag}(m_u,m_d,m_s)$, $Q=e\,{\rm diag}({2\over3},-{1\over3},-{1\over3})$ is the charge matrix, $f$ is the bare pion (or kaon) decay constant, and $C$ is a constant that determines the electromagnetic mass splittings at tree level.
${\cal L}_{\rm gf}$ and ${\cal L}_{\rm ghost}$
are the ghost and gauge-fixing terms in the chiral Lagrangian, but their expressions are not needed in this letter.
Finally, $\Sigma$ is the standard 
parameterization of the Goldstone  fields,
\eq{
\Sigma=\exp\left[i{\lambda_a\phi_a}\over f\right],
}
where
$\lambda_a$ are the Gell-Mann matrices and
$v_{\mu}=v_0\delta_{\mu0}$ with
\eq{
v_0&=
\mbox{$1\over3$}(\mu_B-\mu_S)\mathbb{1}
+\mbox{$1\over2$}\mu_I\lambda_3
+\mbox{$1\over\sqrt{3}$}\mu_S\lambda_8\;,
\label{v000}
}
where the baryon, isospin, and strangeness chemical potentials
expressed in terms of the three quark chemical potentials $\mu_f$
are
\eq{
\mu_B&={3\over2}(\mu_u+\mu_d)\;,\\
\mu_I&=(\mu_u-\mu_d)\;,\\
\mu_S&={1\over2}(\mu_u+\mu_d-2\mu_s)\;.
}
Alternatively, if we
introduce the chemical potentials $\mu_{K^{\pm}}={1\over2}\mu_I+\mu_S$ and $\mu_{K^{0}}=-{1\over2}\mu_I+\mu_S$, we can write
\eq{
v_0&=
\mbox{$1\over3$}(\mu_B-\mu_S)\mathbb{1}
+{1\over2}\mu_{K^{\pm}}\lambda_Q
+{1\over2}\mu_{K^{0}}\lambda_K\;,
\label{mui2}
}
where $\lambda_Q=\lambda_3+{1\over\sqrt{3}}\lambda_8$
and $\lambda_K=-\lambda_3+{1\over\sqrt{3}}\lambda_8$.
Note that $\{\lambda_4,\lambda_5,\lambda_Q\}$
and $\{\lambda_6,\lambda_7, \lambda_K\}$
form the remaining two $SU(2)$ algebras of $SU(3)$.
After having introduced the chemical potentials, the
symmetry is reduced to $U(1)_I\times U(1)_Y$
or equivalently $U(1)_Q\times U(1)_K$.
In the absence of electromagnetic interaction, these are
global symmetries. Including them, the $U(1)_Q$ symmetry
is being gauged and is thus local.
Since the unit operator commutes with all the generators of $SU(3)$,
the results will be independent of $\mu_B$, which simply reflects
that the mesons have zero baryon number and that $\Sigma$
is a singlet under $U(1)_B$ transformations.

By expanding the Lagrangian Eq.~(\ref{lagel}) 
with vanishing chemical potentials
to second order in the fields $\phi_a$, we can read off the
tree-level meson masses in the vacuum.
Considering first the case without EM contributions, we have 
\eq{
m_{\pi^0}^2 &= m_{\pi^{\pm},0}^2 =  B_0 (m_u + m_d) \; ,\\
m_{K^\pm,0}^2 &= B_0(m_u + m_s)\;, \\
m_{K^0{/\bar{K}^0}}^2 &= B_0(m_d + m_s)\;, \\
m_{\eta}^2 &= \frac{1}{3} B_0 (m_u + m_d + 4 m_s)\;.
}
Including EM contributions, the masses of the charged particles become
\eq{
m_{\pi^\pm}^2 &= m_{\pi^\pm,0}^2 + \Delta m_\text{EM}^2\;,\\
m_{K^\pm}^2 &= m_{K^\pm,0}^2 + \Delta m_\text{EM}^2\;,
}
where $\Delta m_{\rm EM}^2=2C {{e^2}\over f^2}$.

In order to understand the phase diagram in three-flavor
QCD, it may be useful to first take a look at the
two-flavor case and pion condensation.
In this case $v_0={1\over3}\mu_B\mathbb{1}+{1\over2}\mu_I\tau_3$,
where $\tau_a$ are the Pauli matrices. For $\mu_I=0$,
the vacuum state $\Sigma_0$ is given by the $2\times2$ unit matrix, $\Sigma_0=\mathbb{1}$.
For finite $\mu_I$, the most general ansatz for the
ground state is~\cite{isostep,isostep2}
\eq{
\Sigma_{\alpha}^{\pi^{\pm}}&=\cos\alpha\mathbb{1}
+i\hat{\phi}_a\tau_a\sin\alpha\;,
\label{ansatz1}
}
where $\hat{\phi}_a$ are variational parameters and $\alpha$
can be thought of as a tilt or rotation angle of the
ground state. The variational parameters satisfy 
$\hat{\phi}_1^2+\hat{\phi}_2^2+\hat{\phi}_3^2=1$ to ensure that the
ground state is normalized, i.e $(\Sigma_{\alpha}^{\pi^{\pm}})^{\dagger}\Sigma_{\alpha}^{\pi^{\pm}}=\mathbb{1}$.
The thermodynamic potential $\Omega_0$ is a function of the
parameters $\hat{\phi}_a$ and the tilt angle $\alpha$.
After a straightforward calculation using the ansatz Eq.~(\ref{ansatz1}), $\Omega_0$ becomes
\eq{
\nonumber
\Omega_0&=
-f^2m_{\pi^0}^2\cos\alpha
\\ & \quad
-{1\over2}f^2
\left(\mu_I^2-\Delta m^2_{\rm EM} \right)\sin^2\alpha(\hat{\phi}_1^2+\hat{\phi}_2^2)
\;.
}
We notice that the energy is minimized for
$\hat{\phi_3}=0$, so that neutral pions do not condense
(the third component of the isospin is zero for a neutral pion).
Thus $(\hat{\phi}_1^2+\hat{\phi}_2^2)=1$ and
the energy is independent of the values of the parameters $\hat{\phi}_1$
and $\hat{\phi}_2$,
this degeneracy reflects
the breaking of the $U(1)_I$ symmetry in the pion-condensed
phase. We can therefore choose $\hat{\phi}_2=1$ without loss
of generality, which yields 
\eq{
\Sigma_{\alpha}^{\pi^{\pm}}&=
\cos\alpha\mathbb{1}+i\tau_2\sin\alpha=
e^{i\alpha\tau_2}\;,\\
\Omega_0&=
-f^2m_{\pi^0}^2\cos\alpha
-{1\over2}f^2
\mu_{I,\rm eff}^2
\sin^2\alpha
\;,
\label{v00}
}
where $\mu_{I,\rm eff}^2=\mu_I^2-\Delta m^2_{\rm EM}$.
The first term arises from the second term in the
Lagrangian~(\ref{lagel}), while the second term comes from 
the second and third term. In minimizing the thermodynamic potential there is a competition
between two terms in Eq.~(\ref{v00}), where the
first term prefers $\alpha=0$ and the second term
$\alpha={\pi\over2}$. The optimum value is found by
minimizing $\Omega_0$, giving
\bqa
\cos\alpha^*=
\left\{ 
\begin{array}{ll}
1\;,  &  \mu_I^2 \leq m_{\pi^{\pm}}^2, \\
{m_{\pi^0}^2} \over {\mu_{I, {\rm eff} }^2 } \;,
&\mu_I^2 \geq m_{\pi^{\pm}}^2 .\\
\end{array}
\right. \;
\eqa
Thus there is a transition from the vacuum phase 
to a pion-condensed phase at $\mu_I^2=m_{\pi^{\pm}}^2$.
In the chiral limit, $\alpha^*={\pi\over2}$
and the quark condensate is rotated into a pion condensate for any nonzero $\mu_I$.

We next consider three flavors. For $\mu_S=0$, the two-flavor
results and the fact that $\{\lambda_1,\lambda_2,\lambda_3\}$
form an $SU(2)$ algebra
suggest that there is a
transition to a pion-condensed phase at  $\mu_I^2=m_{\pi^{\pm}}^2$.
Similarly, for $\mu_K^0=0$, we expect a transition to 
a phase with a charged kaon condensate at $\mu_{K^{\pm}}^2=m_{K^{\pm}}^2$.
Finally, for $\mu_{K^\pm}=0$, a transition to a phase with
a neutral kaon condensate takes place at $\mu_{K^0}^2=m_{K^{0}}^2$.
In the pion-condensed phase the $U(1)_I$ (and therefore also
$U(1)_Q$) is broken. In the kaon condensed phases, either
the $U(1)_Q$ or the $U(1)_K$ symmetry is broken.
Without electromagnetic effects, these phases are superfluid.
There is a single Goldstone boson associated with this
breaking.
If one includes them, the charged pion and kaon condensates
are superconductors since the gauged $U(1)_Q$ is broken.
There is no Goldstone boson in these phases as the Higgs 
mechanisms eats this degree of freedom giving rise to
massive photons with three polarizations.
The form of the ground states is therefore
\eq{
\Sigma_{\alpha}^{\pi^{\pm}}=e^{i\alpha\lambda_2}\;,
\hspace{0.3cm}
\Sigma_{\beta}^{K^{\pm}}=e^{i\beta\lambda_5}\;,
\hspace{0.3cm}
\Sigma_{\gamma}^{K^0/\bar{K}^0}=e^{i\gamma\lambda_7}\;,
}
where $\alpha$, $\beta$,  and $\gamma$ are the tilt angles
in the different cases and the superscript indicates the
condensing meson. Generally, one should allow for the simultaneous condensation
of two or more mesons. However, this is not supported by actual calculations.
The relevant term from the static Lagrangian, 
$-{1\over4}f^2\langle[v_0,\Sigma_{\alpha}][v_0,\Sigma_{\alpha}^{\dagger}]\rangle$ does not
prefer multiple condensates~\cite{martin2}.

In analogy with the two-flavor results,
the transition points between the vacuum and 
Bose-condensed phases are determined by
\eq{
\mu_I^2=m_{\pi^{\pm}}^2\;,
\hspace{0.3cm}
\mu_{K^{\pm}}^2=m_{K^{\pm}}^2\;,
\hspace{0.3cm}
\mu_{K^0}^2=m_{K^0/\bar{K}^0}^2\;.
}
These equations define lines in the $\mu_I$--$\mu_S$ plane
and their intersection define the region of the vacuum phase.
Outside this region, the Bose-condensed phase with the higher pressure wins.

In the pion-condensed phase, the pressure and isospin density
are given by 
\eq{
p&=-\Omega_0\Big|_{\alpha = \alpha^*}=
{1\over2}f^2\mu_{I,\rm eff}^2\left[1-{m_{\pi^{\pm}, 0}^2\over\mu_{I,\rm eff}^2}\right]^2\;,
\label{ppion}
\\
n_I&=-{\partial \Omega_0\over\partial\mu_I}\bigg|_{\alpha = \alpha^*}=
f^2\mu_I\left[1-{m_{\pi^{\pm}, 0}^4\over\mu_{I,\rm eff}^4}\right]\;,
\label{npion}
}
where we here and in the following have subtracted a constant
such that $p=0$ in the vacuum phase.
The strangeness density $n_S=-{\partial \Omega_0\over\partial_{\mu_S}}\big|_{\alpha = \alpha^*}$
vanishes.
As $\mu_I\rightarrow m_{\pi^{\pm}}$ from above, the isospin
density goes to zero continuously, which suggests that the
transition to the vacuum state is of second order.
In order to find the critical exponent, we construct a
Landau-Ginzburg energy functional by expanding the thermodynamic
potential around $\alpha=0$. This yields, after 
omitting a constant term,
\eq{
\nonumber
\Omega_0&=-{1\over2}f^2[\mu_I^2-m_{\pi^{\pm}}^2]\alpha^2
\\ & \quad
+{1\over24}f^2[4\mu_I^2-m_{\pi^{\pm}}^2-3\Delta m_{\rm EM}^2]\alpha^4+\cdots\;.
\label{landau}
}
We define a critical chemical potential $\mu_I^c$ by the
vanishing of the coefficient of the quadratic term.
This yields $\mu_I^c=m_{\pi^{\pm}}$.
Since the quartic term is positive when evaluated at the critical chemical potential,
the transition is of second order as expected. Since  $\alpha^*\propto[\mu_I^2-{\mu_I^c}^2]^{1\over2}$,
for small $\mu_I$,
the mean-field critical exponent is ${1\over2}$.

In the $K^{\pm}$-condensed phase, the isospin and strangeness densities
are given by
\eq{
\label{p2}
p&={1\over2}f^2
\mu_{K^{\pm}{,\rm eff}}^2\left[1-{m_{K^{\pm}{, 0}}^2\over\mu_{K^{\pm},\rm eff}^2}\right]^2\;,\\
n_I&={1\over2}f^2\mu_{K^{\pm}}
\left[1- 
{m_{K^{\pm}{, 0}}^4\over\mu^4_{K^{\pm},\rm eff}}
\right]
\;,\\
n_S&=f^2\mu_{K^{\pm}}\left[1-{m_{K^{\pm}{, 0}}^4\over \mu^4_{K^{\pm},\rm eff}}\right]\;,
\label{ns2}
}
where $\mu^2_{K^{\pm},\rm eff}=\mu^2_{K^{\pm}}-\Delta m_{\rm EM}^2$.
The result for the $K^0/\bar{K}^0$ condensed phases is obtained by making 
the substitutions $\mu^2_{K^{\pm}}\rightarrow \mu^2_{K^0}$,
$\mu^2_{K^{\pm},\rm eff}\rightarrow \mu^2_{K^0}$, and  $m^2_{K^{\pm},0}\rightarrow m^2_{K^0}$
in Eqs.~(\ref{p2})--(\ref{ns2}).

Comparing the pressure of the phases with charged condensates, the transition line, on which
the pressures~(\ref{ppion}) and~(\ref{p2})
are equal, is determined by
\eq{
\nonumber
\mu_{K^{\pm}{\rm,eff}}
&=
\frac{1 }{2\mu_{I{\rm,eff}}}
\bigg(
\mu_{I,{\rm eff}}^2 - m_{\pi^{\pm}{,0}}^2\\
&\quad+
\sqrt{
(\mu_{I,{\rm eff}}^2-m_{\pi^{\pm}{,0}}^2)^2
+4\mu^2
_{I{\rm,eff}} m^2_{K^\pm {, 0}}
}
\bigg)\;.
\label{phaseline}
}
A similar equation for
the transition line between the neutral and charged kaon condensates
can be obtained.

So far we have said nothing about $f$ appearing in the
various expressions for the pressure and the densities.
In an ${\cal O}(p^4)$ analysis of the phase transition, similar
to Eq.~(\ref{landau}), $f$ would be replaced by 
$f_{\pi}$ if the transition is to a pion-condensed phase
and $f_K$ if the transition is to a kaon-condensed phase.
It is therefore natural to identify $f$ with the physical 
value for $f_{\pi}$ or $f_K$ depending on which phase we are
studying. Identifying $f$ with $f_{\pi}$ also guarantees that
the two and three-flavor results for the pion-condensed phase
at ${\cal O}(p^2)$ are the same.
In the next section, we therefore set 
$f=f_{\pi}$, whose experimental value is~\cite{PDG}
\eq{
f_{\pi} &= 92.2,{\rm MeV}\;.
}
In order to generate the phase diagram, we need the physical
values for the pion and kaon masses 
taken from the Particle Data Group~\cite{PDG},
\begin{align*}
m_{\pi^0}&=134.98\,{\rm MeV}\;, 
&
m_{\pi^{\pm}}&=139.57\,{\rm MeV}\;,
\\
m_{K^{\pm}}&=493.68\,{\rm MeV}\;,
&
m_{K^0}&=497.61\,{\rm MeV}
\;.
\end{align*}

\begin{figure}[!htb]
         \centering
         \includegraphics[width=\columnwidth]{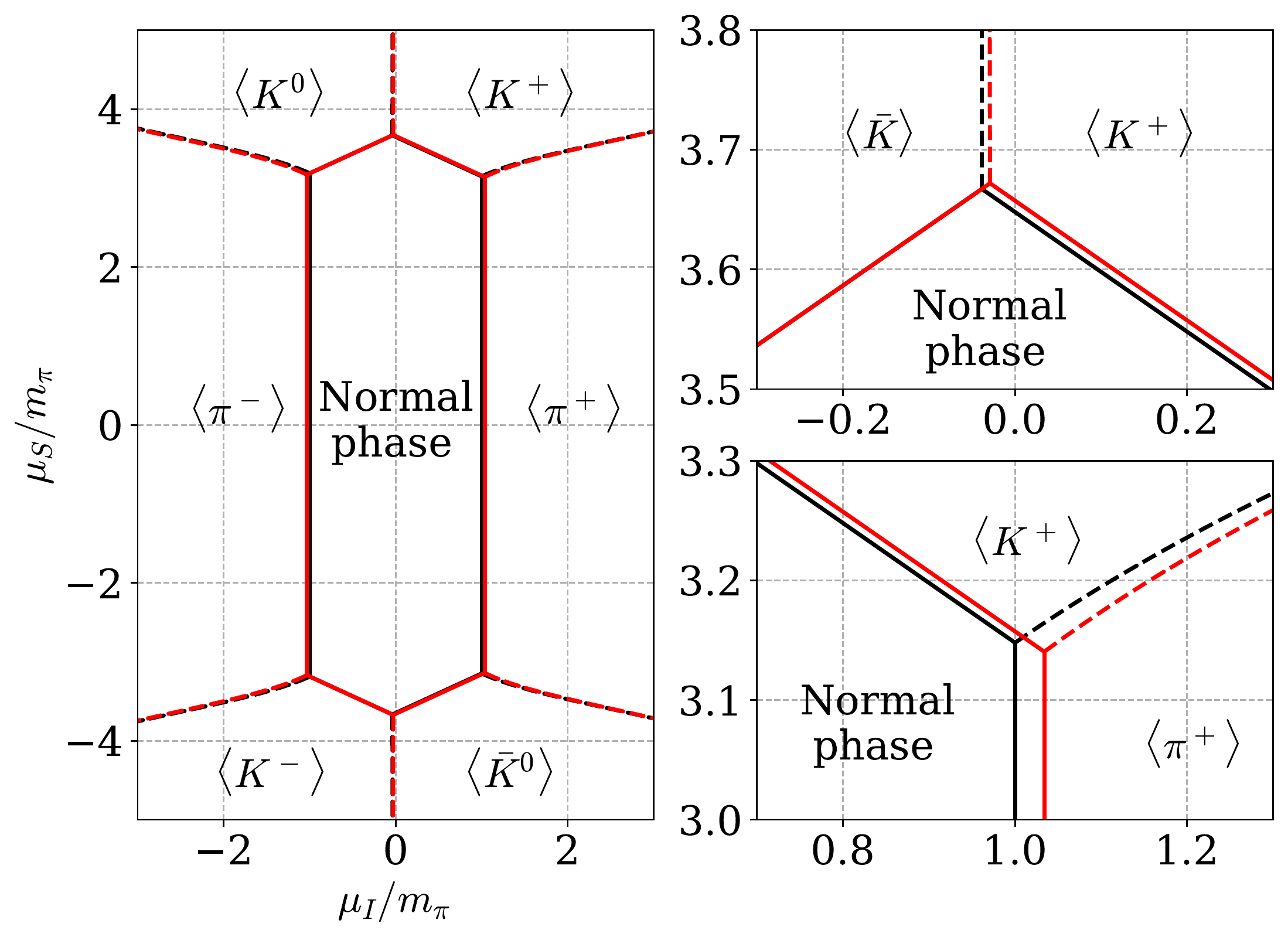}
        \caption{Phase diagram as predicted by $\chi$PT 
        in the $\mu_I$--$\mu_S$-plane. The black lines are results without
electromagnetic interactions, while red lines are results including them. To the right, we have
zoomed in on two of the triple points. Upper panel shows the intersection of the normal, neutral kaon
condensed and charged kaon condensed phases, while the lower shows the intersection of the
normal, pion condensed, and charged kaon condensed phases.}
        \label{diagram}
\end{figure}
In Fig.~\ref{diagram}, we show the phase diagram at $T=0$
in the $\mu_I$--$\mu_S$ plane. 
The black lines are the transition lines without electromagnetic
interactions and the red lines are with electromagnetic interactions. 
The former was first obtained by Kogut and Toublan~\cite{kogut33} with $m_u = m_d$.
The transitions from the
normal phase to a meson-condensed phase is always second order
with mean-field exponents. The transitions are in the 
$O(2)$ universality class, as discussed above. The transitions
between the various condensed phases are always first order
and involves the competition between the order parameters
between the different phases. As we cross the transition lines, the order parameters as well as the densities $n_I$ and $n_S$
jump discontinuously. For example, coming from the $\pi^+$-condensed phase and crossing the the transition line to the
$K^+$-condensed phase, the strangeness density $n_S$ jumps from
zero to the value given by Eq.~(\ref{ns2}).
The small offset of the dashed vertical lines is due to 
the mass difference between the charged and neutral 
kaons, which is both due to $\Delta m_\text{EM}\neq0$, and $m_u\neq m_d$.
These contributions, however, pull in opposite directions, 
as we see in the phase diagram. The contribution due to the difference in
quark masses adds to the mass of the $K^0/\bar{K}^0$ meson, which is why the black transition
line between the kaon condensate is to the left of the $\mu_I = 0$ line, 
while the electromagnetic contribution adds to the mass of the charged kaon,
which is why the red line is between these two lines. 
The partition function in the normal phase is independent of the two chemical 
potentials $\mu_I$ and $\mu_S$, which is the Silver Blaze
property~\cite{cohen0}. It implies that all the thermodynamic functions are
constant, for example that the pressure and the
charge densities vanish. Similarly, the
partition function in  the
condensed phases is independent of 
one of the chemical potential $\mu_I$ or $\pm{1\over2}\mu_I+\mu_S$,
implying that $n_S=0$, $n_{K^{\pm}}=0$, or $n_{K^0/\bar{K}^0}=0$, respectively. 

The phase diagram has been mapped out under the assumption that the chemical potentials
are small so that $\chi$PT is inside its region of validity. 
The expansion parameter is $M/4\pi f$, where $M$ is a mass
or a chemical potential. An expansion parameter of ${1\over2}$
corresponds to $M\approx 565$ MeV.
Another constraint on the region of validity of $\chi$PT is the appearance of new resonances not in the model. The first new resonance to appear is the $\rho$-meson, at $M \approx 770$ MeV.
A natural question to ask
is: what happens at larger values of $\mu_I$ and $\mu_S$. For asymptotically large
values of $\mu_I$, asymptotic freedom implies that quarks are weakly interacting and
a description in terms of these degrees of freedom applies~\cite{isostep,isostep2}. 
One-gluon exchange gives rise to an attractive channel, rendering the Fermi surface unstable
to the formation of Cooper pairs of $\bar{u}$ and $d$ quarks or 
$\bar{d}$ and $u$ depending on the sign of $\mu_I$. 
The resulting BCS phase is characterized by the same order parameter as the pion-condensed
phase~\cite{isostep,isostep2}.

\section{Pion stars}
\label{stars}
Describing a pion star using the perfect fluid approximation as done for neutron stars,
requires an equation of state. For an ideal Bose gas at zero temperature, all the
bosons are in the zero-momentum state, the pressure is vanishing,
and there is no equation of state~\cite{bosonstars}.
In contrast to neutron stars, where the Pauli principle gives rise to a quantum pressure
at $T=0$, we need interactions in the bosonic case. 
In the pion-condensed phase, the mean-field pressure is given in Eq.~(\ref{ppion}),
which vanishes for $f=f_{\pi}=0$.
Using $p$ and $\epsilon$ to describe matter, the  
Tolman-Oppenheimer-Volkov (TOV) equation for a 
spherically symmetric star follows directly from
Einstein's fields equation.
The continuity equation and the 
TOV equation for a spherically symmetric star are
\eq{
{dm\over dr}&=4\pi r^2\epsilon(r)\;,
\label{dmdr}
\\
\nonumber
{dp\over dr}&=-{{G\epsilon(r)m(r)}\over r^2}
\left[1+{p(r)\over \epsilon(r)}\right]
\\ &\quad\times
\left[1+{4\pi r^2p(r)\over m(r)}\right]
\left[1-{2Gm(r)\over r^2}\right]^{-1}\;.
\label{tov1}
}
The TOV equation is the general-relativistic generalization of the corresponding equation
in Newtonian physics, which is derived from imposing hydrostatic equilibrium throughout the star.
The two equations above are coupled and involve three unknowns, namely the pressure $p(r)$, the energy density
$\epsilon(r)$, and the gravitational mass inside a shell of radius 
$r$, $m(r)$. We therefore need the
equation of state $p(\epsilon)$ to close the system.
The pressure is a function of one or more chemical potentials depending on the composition of the
star. Imposing electric charge neutrality and assuming chemical equilibrium yield a number
of a constraints among them and one ends up with a single free chemical potential, for example
$\mu_I$. The pressure and the energy density can then be parameterized uniquely by it.

The energy density is given by a Legendre transform of the pressure,
\eq{
\epsilon&=-p+\sum_in_i\mu_i\;,
}
where the sum $i$ is over all relevant chemical potentials.

\begin{figure}[htb]
\centering  
\includegraphics[width=\columnwidth]{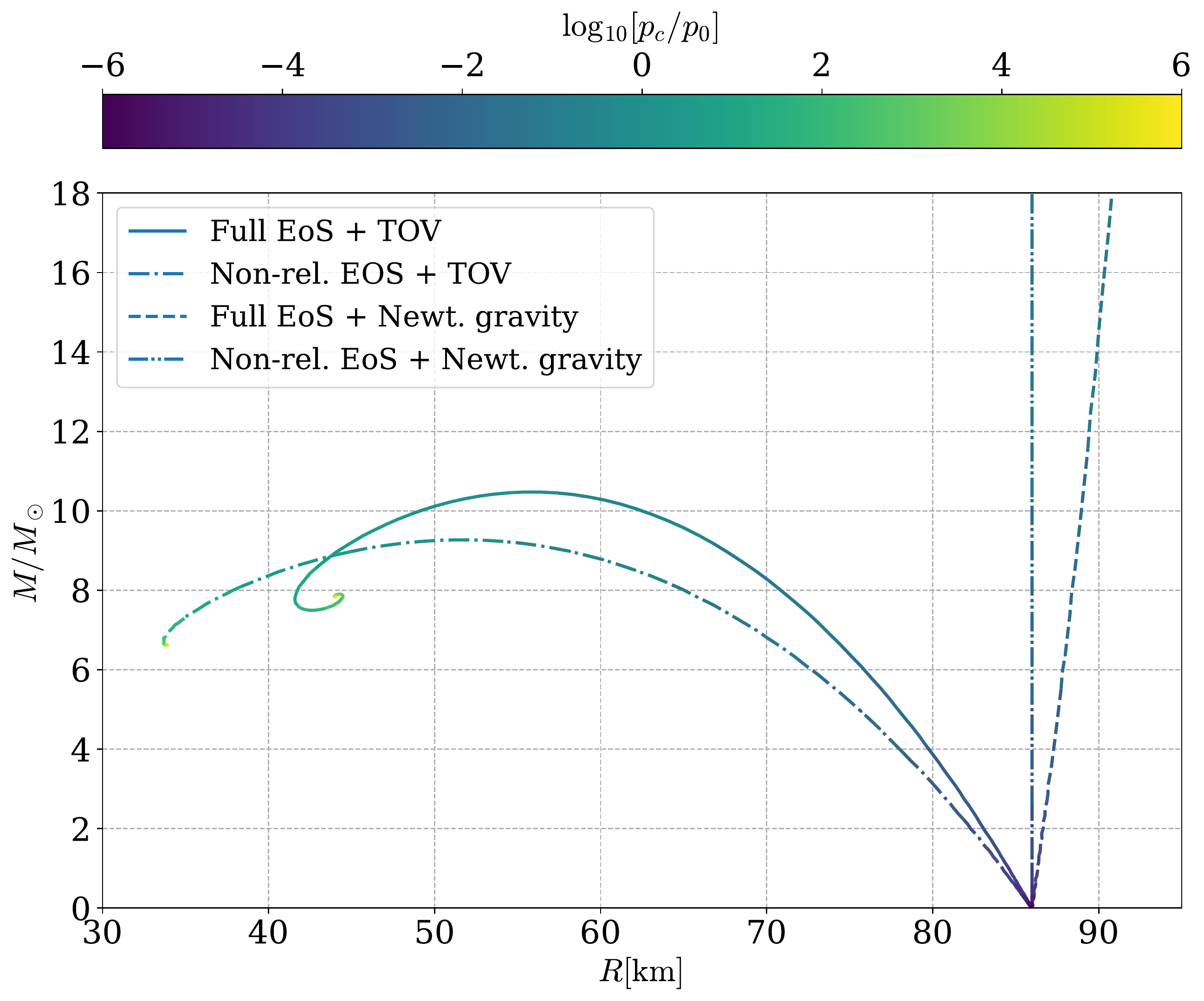}
\caption{Mass-radius relation for a pure pion star
without electromagnetic interactions
in different approximations. See main text
for details.}%
\label{pionrel0}
\end{figure}

We first discuss a pure pion star with the thermodynamic quantities at ${\cal O}(p^2)$
and set for simplicity
$\Delta m^2_{\rm EM}=0$ in Eqs.~(\ref{eos})--(\ref{expe}) below.
For the pion-condensed phase, we can find the energy density using Eqs.~(\ref{ppion})
and~(\ref{npion}). Expressing $\epsilon$ as a function of $p$,
we obtain the EoS
\eq{
\epsilon(p)&=-p+2\sqrt{p(p+2f_{\pi}^2m_{\pi^\pm,0}^2)}\;.
\label{eos}
}
In the ultrarelativistic limit $\mu_I\rightarrow\infty$,
the EoS reduces to $\epsilon=p$, which
is a polytrope with index $n=\infty$ or $\gamma=1$.
In the opposite limit, we write the isospin chemical potential as 
$\mu_I=m_{\pi^\pm{,0}}+\mu_{\rm NR}$, where $\mu_{\rm NR}\ll m_{\pi^\pm{,0}}$
is the non-relativistic chemical potential. One can then show that
\eq{
p&=2f_{\pi}^2\mu_{\rm NR}^2\;,\\
n_I&=4f_{\pi}^2\mu_{\rm NR}
\;,\\
\label{expe}
\epsilon&=m_{\pi^{\pm}{,0}}n_I
+{1\over8f_{\pi}^2}
n_I^2\;.
}
This yields the
equation of state is $p=K\epsilon^2$, i.e. it is a polytrope
with polytropic index $n=1$ and constant $K^{-1} = 8 m_{\pi^\pm}^2f_\pi^2$. It is also a polytrope with
$n=1$ for $\Delta m_{\rm EM}^2\neq 0$, but now with
$K^{-1}=8 m_{\pi^\pm}^4f_\pi^2/m_{\pi^0}^2$.
The pressure and energy density
are those of a weakly interacting nonrelativistic Bose gas 
with an $s$-wave scattering length $a={m\over16\pi f_{\pi}^2}$ in the mean-field approximation, first obtained by 
Bogoliubov~\cite{bogo}.~\footnote{To obtain the energy density of a dilute Bose gas, we
subtract the leading term $m_{\pi^{\pm},0}n_I$ associated with the rest mass energy.}
For stars with a sufficiently low central pressure, we expect that
the EoS of a dilute Bose gas is a good approximation.
Moreover, if $\epsilon(r)\gg p(r)$ and $r \gg 2Gm(r)$, the 
TOV equation reduces to the Newtonian equation for hydrostatic
equilibrium, namely the Lane-Emden equation. It is expected
that this is also a good approximation for sufficiently low-mass
stars. Solving the Lane-Emden equation for $\gamma=2$, one 
finds that the radius $R$ is independent of its mass, i.e. it
corresponds to a vertical line in the mass-radius diagram.~\footnote{This limit is analogous to the
Chandrasekhar limit for white dwarfs, where its radius is
independent if its mass attains its maximum value, approximately
$1.4M_{\odot}$.}
The radius in the Newtonian limit is given by
\eq{
\label{limit radius}
R &= \pi  \sqrt{ \frac{K}{2 \pi G} }.
}
This is shown in Fig.~\ref{pionrel0}, where we plot the
mass-radius relation in different approximations
for a pure pion star without electromagnetic interactions.

\begin{figure}[htb]
\centering  
\includegraphics[width=\columnwidth]{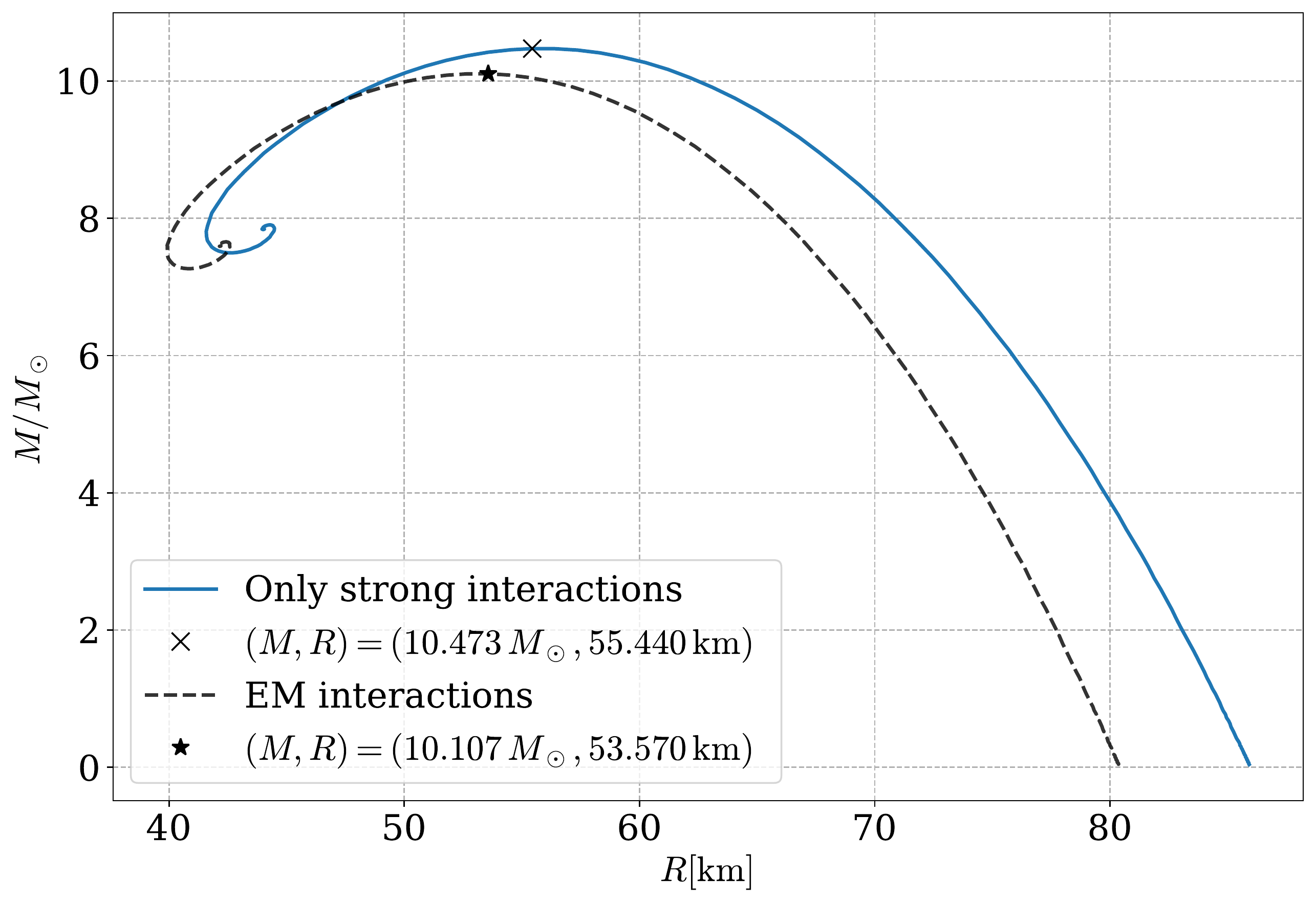}
\caption{Mass-radius relation for a pure pion star. See main text
for details.}%
\label{pionrel}
\end{figure}
In Fig.~\ref{pionrel}, we show the mass-radius relation for a pure pion star. 
The solution to the TOV equation without electromagnetic interactions 
is given by the solid line, where the stable stars
are to right of the maximum mass indicated by a cross.
The maximum mass is $M_{\rm max}=10.47M_{\odot}$ and the corresponding
radius is $R=55.4$ km. 
The mass-radius relation
including electromagnetic interactions is given by the black dashed line in Fig.~\ref{pionrel}. The maximum mass is now 
$10.1M_{\odot}$ with a radius of $53.6$ km.
The maximum radius for vanishing mass is $R_{\rm max}=86$ km.
If we include electromagnetic effects, the maximum radius
is reduced to $R_{\rm max}=80$ km.
These numerical results are in excellent agreement with the limit Eq.~(\ref{limit radius}).

We next discuss electric charge neutrality. Since condensed pions
are electrically charged, a pure pion star has a macroscopic electric charge.
However, due to the Coulomb repulsion among
the pions, there is an enormous energy cost of having
bulk matter that is not electrically neutral~\cite{densestars}.
We will therefore impose local electric charge neutrality by adding a background charge
of leptons. This amounts to requiring
\eq{
n_I-\sum_ln_l&=0\;,
\label{nq}
}
where the sum is over lepton species $l$.
Chiral perturbation theory can be extended to include the light leptons as dynamical degrees of
freedom~\cite{leptons}. The ${\cal O}(p^2)$-term one should add to the Lagrangian Eq.~(\ref{lagel}) is
\eq{
\nonumber
{\cal L}_2=
\sum_l&
\left[\bar{l}(i\partial\!\!\!/-\mu_l\gamma_0+eA\!\!\!/-m_l)l
\right. \\ &\left.
+\bar{\nu}_{lL}
(i\partial\!\!\!/-\mu_{\nu_{l}}\gamma_0)\nu_{lL}\right]\;,
}
where we include the electron and the muon in the sum,
$m_l$ is the lepton mass, $\mu_l$ and $\mu_{\nu_{l}}$ are the corresponding chemical potentials.
The neutrinos are massless, $m_{l_{\nu}}=0$, and the subscript $L$
indicates that there are only left-handed neutrinos.
In the power counting scheme adopted in Ref.~\cite{leptons}, a fermion bilinear and 
lepton mass both count as ${\cal O}(p)$.
Below we include the leading contributions to the pressure and energy density, which according
to the above counting scheme are ${\cal O}(p^4)$.
It is therefore not consistent to include strong and electromagnetic interactions among the pions to  ${\cal O}(p^2)$  and the leptons to ${\cal O}(p^4)$. We therefore 
include the ${\cal O}(p^4)$ corrections from the pions in the pressure, isospin density,
and energy density. Details of the calculations can be found in Ref.~\cite{martin2}
or in a forthcoming paper~\cite{martin3}.
Moreover, the ${\cal O}(p^4)$ electromagnetic effects
are prohibitively difficult to calculate, but since we are mainly
interested in the comparison with the lattice simulations, we
do not need them anyway.

A lepton with chemical potential $\mu_l>m_l$
contributes to the pressure and energy density as
\eq{
p_l&=
{2m_l^4\over3(4\pi)^2}
\left[
(2x_l^3-3x_l)\sqrt{1+x_l^2}+3{\rm arcsinh}\,x_l
\right]\;,
\\
\epsilon_l&=
{2m_l^4\over(4\pi)^2}
\left[
(2x_l^3+x_l)\sqrt{1+x_l^2}-{\rm arcsinh}\,x_l
\right]\;,
\label{cont2}
}
where $x_l={\sqrt{\mu_l^2-m_l^2}\over m_l}$ is 
the dimensionless ratio of the Fermi momentum and the
mass of the lepton. 
The contribution to the charge density from a lepton is
\eq{
n_l&={\partial p\over\partial\mu_l}
={16\over3(4\pi)^2}(\mu_l^2-m_l^2)^{3\over2}\;.
}
Considering again the nonrelativistic limit for a fermion, 
the standard results for the lepton pressure and energy density are
\eq{
p_l&={16m_l^4\over15(4\pi)^2}(\mu_l^2-m_l^2)^{5\over2}
\;,
\hspace{0.5cm}
\epsilon_l=m_{\pi^{\pm}}n_I\;,
}
which leads to the equation of state $p_l=K\epsilon_l^{5\over3}$.
The neutrinos are massless so
$p$ and $\epsilon$ are given by the their ultrarelativistic limit,
including a factor of ${1\over2}$ due to the absence of right-handed
neutrinos and left-handed antineutrinos,
\eq{
p_{\nu_l}=
{2\mu_{\nu_l}^4\over3(4\pi)^2}\;,\hspace{1cm}
\epsilon_{\nu_l}&={2\mu_{\nu_l}^4\over(4\pi)^2}\;.
\label{urfermion}
}

Pions are unstable particles that mainly decay via weak interactions.
The dominant decay mode is $\pi\rightarrow\mu+\nu_{\mu}$.
In chemical equilibrium, the reactions rates of
$\pi^+\rightarrow l^++\nu_{\mu}$
and $\pi^-\rightarrow l^-+\bar{\nu}_{\mu}$
are the same, implying the relation
\eq{
\mu_I&=\mu_l-\mu_{\nu_l}\;,
\label{chem}
}
where $l$ is either an electron or a muon.
Although we ignore the neutrino masses in the analysis, the
nonzero masses imply a relation between their chemical potentials
via neutrino oscillations. Chemical equilibrium among the two species yields
\eq{
\mu_{\nu_e}&=\mu_{\nu_{\mu}}\;,
}
or equivalently $\mu_e=\mu_{\mu}$. We have six chemical potentials, $\mu_I$, $\mu_Q$, 
$\mu_e$, $\mu_{\mu}$, $\mu_{\nu_e}$, and $\mu_{\nu_{\mu}}$. Chemical equilibrium and 
charge neutrality reduce this number to a single independent chemical potential.

We note that at the transition when $\mu_I=m_{\pi^{\pm}}$, the lepton density also vanishes
due to the charge neutrality constraint (\ref{nq}). However, chemical equilibrium,
Eq.~(\ref{chem}), implies that the neutrino chemical potential and consequently the
neutrino density is nonzero at the transition point, i.e. on the surface of the pion star.
The star therefore has a neutrino atmosphere with an EoS $\epsilon=3p$.

\begin{figure}[htb!]
\centering
\includegraphics[width=\columnwidth]{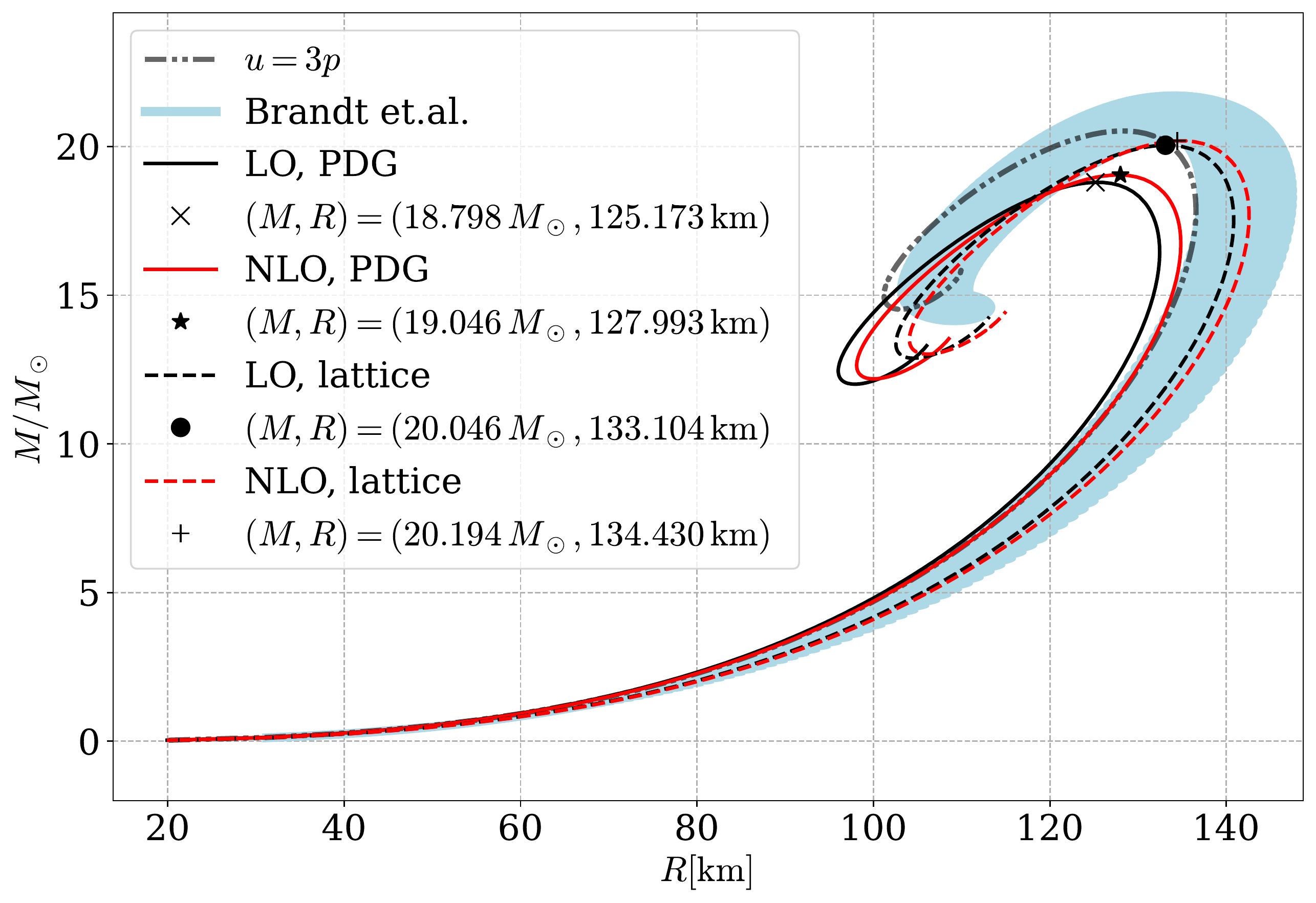}
\caption{Mass-radius relations for a pion star for the different
cases as indicated by the legends. Electromagnetic interaction are not included.
The dashed lines are results using the EoS obtained from lattice calculations~\cite{gergy3}.}%
\label{pionerel2}
\end{figure}

In the absence of electromagnetic interactions, we take the physical mass of the neutral pion and kaon as our common meson masses.
The masses of the leptons are also taken from Ref.~\cite{PDG}
\begin{align}
m_e &= 0.51\;{\rm MeV}
\;, &
m_\mu& = 105.6\;{\rm MeV}\;.
\end{align}
In order to compare with lattice simulations, we also use the set
in Ref.~\cite{gergostar}.
The central values are given by 
\eq{
m_{\pi}=131\,{\rm MeV}\;,\;
m_{K}=481\,{\rm MeV}\;,\;
f_{\pi}=90.8\,{\rm MeV}\;.
}
In Fig.~\ref{pionerel2}, we show the mass-radius for pion stars including 
leptons and neutrinos. The blue band is obtained from lattice simulations including the statistical and systematic errors in the EoS.
The black solid and dashed lines are obtained from the ${\cal O}(p^2)$ EoS from
$\chi$PT using the PDG and lattice values for the meson masses, respectively.
The red solid and dashed lines are obtained from the ${\cal O}(p^4)$
EoS and the same sets of meson masses.
The results show that the mass-radius relations are rather sensitive to the
masses and decay constant. 
This change can mostly be understood as a scaling of the variables in the TOV equation.
As we will discuss further, the EoS is dominated by the neutrinos, so the most important mass scale is 
$p_\text{min} \propto f_\pi^2 m_\pi^2$, where $p_{\rm min}$
is the pressure on the surface of the star.
Scaling the pressure and energy density in the TOV equation by $t^2$ leads to a scaling of $M, R$ 
by $t^{-1}$, so we expect both mass and radius to be proportional to $1 / m_\pi f_\pi$.
Using the lattice constants, this quantity is around $5\%$ larger than using the PDG constants, which accounts for most of the difference in \ref{pionrel}.

Moreover, the results for the
${\cal O}(p^2)$ and ${\cal O}(p^4)$ curves are very close.
The quantity ${\mu_{\pi}\over m_{\pi}}-1$ is equal to $2.0\times10^{-2}$
at ${\cal O}(p^2)$ and $1.8\times10^{-2}$ at ${\cal O}(p^4)$ 
at the center for the heaviest star, denoted by the crosses.
This corresponds to the largest values of $\mu_I$ on the entire
branch. In this region the corrections to the thermodynamic quantities
are small suggesting that the we can reliably use $\chi$PT to calculate the properties of pion stars. 
As noted above, at the surface of the star, $\epsilon=3p$.
Chemical equilibrium and charge neutrality imply that
$\mu_e=\mu_{\mu}=m_e$ and therefore 
$\mu_{\mu_e}=\mu_{\mu_{\nu}}=m_{\pi^{\pm{,0}}}+m_e$.
This yields the pressure on the surface
$p_{\rm min}={4\over3(4\pi)^2}(m_{\pi^{\pm}{,0}}+m_e)^4$.

The equation of state in the charge neutral pion condensate is well approximated by $\epsilon = 3 p$, especially for low pressures, as the neutrino contribution dominates.
In Fig~\ref{pionerel2}, we have included the results of solving the TOV equation with this EoS, and a surface pressure $p = p_{\rm min}$.
The result is the grey dashed curve
in Fig.~\ref{pionerel2}, demonstrating that the 
EoS is dominated by the neutrinos if the star is not too compact.

A further investigation of pion stars requires a more complete understanding of their evolution and instabilities.
A necessary requirement for the stability of stars against spherical perturbations is the increase of their mass as the central pressure increases \cite{glendenning},
\beq
    {d M \over d p_c} \geq 0\;.
\eeq
Brandt et al. showed, as expected, that pion stars including leptons and neutrinos are stable for central pressures below that of the maximum mass configuration~\cite{gergostar}.
In addition to perturbations, an important component of the stability of pion stars is the decay rate of its constituent particles.
As the pion star is modeled at zero temperature, the decay of the pion condensate into neutrinos and charged leptons will be Pauli blocked.
The neutrino atmosphere, however, is not gravitationally bound as we consider neutrinos massless.
This will lead to the depletion of the lepton number at the edge of the star.





\section{Acknowledgements}
The authors would like to thank B. Brandt, G. Endr\H{o}di, and
S. Schmalzbauer for useful discussions as well as their
lattice data for the mass-radius relation of pion 
stars~\cite{priv}.

\end{document}